\documentclass[12pt]{article}

\ifx\pdfoutput\undefined
\usepackage[dvips,bookmarks]{hyperref}
\else
\usepackage{hyperref}
\fi
\hypersetup{colorlinks=false,bookmarksopen,bookmarksnumbered,citecolor=blue,
   pdfstartview=FitH}

\usepackage{latexsym}
\usepackage{amssymb,amsfonts,amsmath}
\usepackage{graphicx} 
\usepackage{indentfirst}
\usepackage{bbm}
\usepackage{amssymb}
\usepackage{verbatim}
\usepackage{amsmath, amsthm,amssymb}
\usepackage{mathrsfs}
\usepackage{hyperref}
\usepackage{amsfonts}
\usepackage{dsfont}
\usepackage{slashed, tensor}
\usepackage{booktabs}
\usepackage{graphicx}
\usepackage{mathrsfs}

\usepackage[colorinlistoftodos,prependcaption,textsize=small]{todonotes}
\parskip=\medskipamount

\newcommand {\cL}{{\cal L}}

\newcommand {\cN}{{\cal N}}
\newcommand {\cO}{{\cal O}}

\def\a{\alpha}
\def\b{\beta}

\def\d{\delta}

\def\g{\gamma}

\def\l{\lambda}

\def\r{\rho}

\def\t{\tau}

\def\x{\xi}
\def\z{\zeta}

\def\L{\Lambda}
\def\O{\Omega}

\def\ri{{\rm i}}

\newcommand{\ve}{\varepsilon}                            

\newcommand{\hf}{\frac12}

\newcommand{\be}{\begin{equation}}
\newcommand{\ee}{\end{equation}}
\newcommand{\bea}{\begin{eqnarray}}
\newcommand{\eea}{\end{eqnarray}}
\newcommand{\non}{\nonumber}
\newcommand{\ba}{\begin{array}}
\newcommand{\ea}{\end{array}}

\newcommand{\bm}[1]{\mbox{\boldmath$#1$}}

\def\double #1{#1{\hbox{\kern-2pt $#1$}}}

\newcommand{\bsubeq}{\begin{subequations}}
\newcommand{\esubeq}{\end{subequations}}

\newcommand{\eps}{{\ve}}

\newcommand{\rd}{\mathrm d}
\numberwithin{equation}{section}

\newcommand{\Tmult}{T}

\begin{document}

\begin{titlepage}
\begin{flushright}
July, 2017 \\
\end{flushright}
\vspace{5mm}

\begin{center}
{\Large \bf New superconformal multiplets and 
 higher derivative invariants in six dimensions 
}
\\ 
\end{center}

\begin{center}

{\bf
Sergei M. Kuzenko${}^{a}$, Joseph Novak${}^{b}$ and Stefan Theisen${}^{b}$
} \\
\vspace{5mm}

\footnotesize{
${}^{a}${\it School of Physics and Astrophysics M013, The University of Western Australia\\
35 Stirling Highway, Crawley W.A. 6009, Australia}}  
~\\
\vspace{2mm}
\footnotesize{
${}^{b}${\it Max-Planck-Institut f\"ur Gravitationsphysik, Albert-Einstein-Institut\\
Am M\"uhlenberg 1, D-14476 Golm, Germany}
}
~\\
\vspace{2mm}

\texttt{sergei.kuzenko@uwa.edu.au, joseph.novak@aei.mpg.de,  stefan.theisen@aei.mpg.de}\\
\vspace{2mm}

\end{center}

\begin{abstract}
Within the framework of six-dimensional ${\cal N}=(1,0)$ conformal supergravity, 
we introduce new off-shell multiplets ${\cal O}{}^{*}(n)$, where $n=3,4,\dots,$ 
and use them to construct higher-rank extensions of the linear multiplet action. 
The ${\cal O}{}^{*}(n)$ multiplets may be viewed as being dual to well-known 
 superconformal ${\cal O}(n)$ multiplets.
We provide prepotential formulations for the ${\cal O}(n)$ and ${\cal O}{}^{*}(n)$ multiplets 
coupled to conformal supergravity. For every ${\cal O}{}^{*}(n)$ multiplet, we construct a higher derivative
invariant which is superconformal on arbitrary superconformally flat backgrounds. 
We also show how 
our results can be used to construct new higher derivative actions in supergravity.

\end{abstract}

\vfill

\vfill
\end{titlepage}

\newpage
\renewcommand{\thefootnote}{\arabic{footnote}}
\setcounter{footnote}{0}


\allowdisplaybreaks


\section{Introduction}

Recently, there has been some interest in higher derivative superconformal invariants
in six dimensions  (6D) \cite{Beccaria:2015uta,Beccaria:2015ypa,Beccaria:2017dmw},
e.g. in the context of general relations between the conformal and chiral anomaly coefficients in 6D $\cN=(1,0)$ 
superconformal theories.  One  
example of such higher derivative models is the $\cN=(1,0)$ superconformal 
vector multiplet theory.  It was formulated in Minkowski superspace
in \cite{ISZ05}, and coupled to conformal supergravity in \cite{BKNT}.\footnote{Its component Lagrangian in 
conformal supergravity was worked out 
in \cite{BNT-M17}.}
In this paper, we propose new off-shell superconformal  
multiplets and use them to construct higher derivative invariants that 
are quadratic in the fields of the new multiplets and naturally generalise the superconformal vector multiplet action.

We recall that the superconformal vector multiplet action 
is the supersymmetric extension of an $F \Box F$ term, where 
$F$ is the two-form field strength of a gauge one-form.  
Its construction starts with the supersymmetric $BF$ invariant\footnote{The notation refers to the six-form 
which is the product of the field strength $F$ and the four-form potential $B$
contained in the linear multiplet. It is also known as the linear multiplet action.}
which is, schematically, the product 
of a vector multiplet with a linear multiplet. It may be 
written as a full superspace integral in $\cN = (1, 0)$ Minkowski superspace 
as follows\footnote{The second form of the invariant in \eqref{BFinv} follows 
from the expressions for the covariant field strengths $W^{\a i}$ and $L^{ij}$ 
in terms of their prepotentials.}
\be I = \int \rd^{6|8} z\, U_{ij} \, {L}^{ij} = \int \rd^{6|8} z\, W^{\a i} \r_{\a i} \ ,
\label{BFinv}
\ee
where the superfield $L^{ij} = L^{(ij)}$ satisfies the defining differential constraint 
for the linear multiplet \cite{BSohnius,HST83},
\be D_\a^{(i} L^{jk)} = 0 \ .
\ee
The superfield $W^{\a i}$ satisfies the differential constraints
\be D_\a^{(i} W^{\b j)} = \frac{1}{4} \d_\a^\b D_\g^{(i} W^{\g j)} \ , \quad D_{\a i} W^{\a i} = 0
\ee
appropriate for a vector multiplet, see  \cite{HST83, Koller83} and references therein. 
The real iso-triplet  $U_{ij}$ in \eqref{BFinv}
is the 6D counterpart of the Mezincescu prepotential
\cite{Mezincescu} for the vector multiplet, 
and $\r_{\a i}$ is the prepotential \cite{GL85} for the linear multiplet. 
The prepotentials  $\r_{\a i}$ and $U_{i j}$  play the role of gauge fields, while  
 $L^{ij}$ and $W^{\a i}$ are the gauge-invariant field strengths of the linear and vector multiplets, respectively. 
One can construct the supersymmetric $F \Box F$ action by building a linear multiplet
from the fields of the vector multiplet. The appropriate composite linear multiplet is
\be
{\bm L}^{ij} = 
\ri \Box D_\g^{(i} W^{\g j)} \ , \quad \Box := \partial^a \partial_a \ .
\label{1.4}
\ee
The $F \Box F$ action is the $BF$ invariant \eqref{BFinv} 
with the replacement $L^{ij} \rightarrow {\bm L}^{ij}$:
\be S_{F \Box F} = \int \rd^{6|8} z \, U_{ij} {\bm L}^{ij} \ .
\ee
The above action turns out to be superconformal\footnote{Unlike 
$D_\g^{(i} W^{\g j)}_{\phantom{\g}} $, the 
higher-derivative linear multiplet 
\eqref{1.4} is a primary superfield.}
and equivalent to the harmonic superspace action given in \cite{ISZ05}. 

In order to formulate generalisations of the above construction, we will consider
$\cO(n)$ multiplets as higher-rank cousins of the linear multiplet, $\cO(2)$, 
and uncover a new family of multiplets 
akin to the vector multiplet. These multiplets will allow us to generalise the $BF$ action in such 
a way that it is now, schematically, the product of the $\cO(n)$ multiplet with the new multiplet. Using 
this invariant we will show how to generate a whole family of superconformal higher derivative invariants
bilinear in the new multiplets.

In what follows, we will use the superspace formulation for conformal supergravity in \cite{BKNT} and 
refer the reader there  for our notation and conventions. The results in 
Minkowski superspace are obtained by  setting the super-Weyl tensor $W^{\a\b}$ to zero  
and replacing the conformal superspace covariant derivatives with those of 
Minkowski superspace, $\nabla_A = (\nabla_a , \nabla_\a^i) \rightarrow (\partial_a , D_\a^i)$. 
In this paper, all (super)fields are chosen to transform in irreducible representations
of the $R$-symmetry group $\rm SU(2)$ so that 
they are 
symmetric 
in their $\rm SU(2)$ spinor indices, for instance
$L^{i_1 \cdots i_n} = L^{(i_1 \cdots i_n)}$.


\section{The $\cO(n)$ multiplets and their prepotentials}

Given an integer $n \geq 1 $, the $\cO(n)$ multiplet is  a primary superfield $L^{i_1 \cdots i_n} 
$ of dimension $2 n$ that 
satisfies the differential constraint\footnote{The $\cO(n)$ multiplets are well known in the literature on supersymmetric field 
theories with eight supercharges in diverse dimensions. 
For 4D $\cN=2$ Poincar\'e supersymmetry, 
general $\cO(n)$ multiplets, with $n>2$, 
were introduced in \cite{KLT,GIO,LR}.
The case $n=4$ was first studied in \cite{SSW}.
The terminology ``$\cO(n)$ multiplet''
was coined in \cite{G-RRWLvU}.
As 6D $\cN=(1,0)$ superconformal 
multiplets, their complete description was given in \cite{LT-M12}
following the earlier approaches in four and five dimensions
\cite{K06,K07,KT-M08,KLRT-M}. }
\be \nabla_\a^{(i_1} L^{i_2 \cdots i_{n+1})}_{\phantom{\a}} = 0 \ . \label{O(n)DefiningConstraint}
\ee
The contraint \eqref{O(n)DefiningConstraint} is primary since the $S$-supersymmetry generator 
$S^\b_j$ annihilates the left hand side (without requiring the constraint). This 
condition actually fixes the dimension of $L^{i_1 \cdots i_n}$. 
The complex conjugate of $L^{i_1 \dots i_{n}}$, which is  defined by
\bea
 \overline{L}_{i_1 \dots i_{n}}
 := \overline{ L^{i_1 \dots i_{n}} } ~,
 \eea
 is also an $\cO(n)$ multiplet.
If $n$ is even, $n=2m$, one can consistently define real $\cO(2m)$ multiplets by 
 imposing the condition $\overline{L}_{i_1 \dots i_{2m}}
 = \ve_{i_1 j_i} \dots \ve_{i_{2m} j_{2m}} 
 L^{j_1 \dots j_{2m}} \equiv L_{i_1 \dots i_{2m}}$.

The $n = 1$ case corresponds to the 6D  version of the Fayet-Sohnius hypermultiplet \cite{Sohnius}, 
which is necessarily on-shell in six dimensions 
(the constraint $ \nabla_\a^{(i} L{}^{j)}_{\phantom{\a}}=0$ implies $\Box L^i =0$)
and, therefore, it will not be considered in what follows.
The case $n = 2$ 
is the linear multiplet, which can be formulated in conformal superspace 
in terms of an unconstrained prepotential $\r_\a^i$ as follows\footnote{The 
derivation of \eqref{2.2} is rather technical  and will be given elsewhere.}
\be L^{ij} = \nabla^{(ijkl} \nabla^\a_{klp} \r_\a^{p)} - \frac{12 \ri}{5} \nabla^{ijkl} \big( \nabla_{\a k} \nabla^{\a\b} \r_{\b l} 
+ 4 \nabla_{\a k} (W^{\a\b} \r_{\b l}) \big) \ ,
\label{2.2}
\ee 
where we have introduced $\nabla^{\a ijk} := \frac{1}{3!} \eps^{\a\b\g\d} \nabla_\b^{(i} \nabla_\g^{j} \nabla_\d^{k)}$
and the covariant projection operator\footnote{In SU(2) superspace, 
it coincides with the one in \cite{LT-M12}.}
\be \nabla^{ijkl} := - \frac{1}{96} \eps^{\a\b\g\d} \nabla_\a^{(i} \nabla_\b^{j} \nabla_\g^k \nabla_\d^{l)} \ .
\ee
The prepotential  is defined up to the following gauge transformations
\be
\d \r_\a^i = \nabla_\a^i \t + \nabla_{\b j} \t_\a{}^\b{}^{ij} \ , \quad \t_\a{}^\a{}^{ij} = 0 \ , 
\label{RhoPrepotGaugeTrans}
\ee
where $\t$ and $\t_\a{}^\b{}^{ij}$ are dimensionless primary superfields.
These gauge transformations leave  $L^{ij}$ invariant.
In the flat-superspace limit, eq. \eqref{2.2} reduces to the one 
given in  \cite{GL85}. 

The field content of the $\cO(n)$ multiplets follows 
from taking spinor covariant derivatives of the superfield $L^{i_1 \cdots i_n}$ 
and analysing the consequences of the constraint \eqref{O(n)DefiningConstraint}. The 
independent component fields are summarised in Table \ref{O(n)fieldContent} 
from which it can be seen that the number of off-shell degrees of freedom 
is $8(n-1) + 8(n-1)$. The cases  $n \leq 3$ 
are special since the component fields for which the number of SU(2) indices become negative are 
truncated away. Note that, as the Dirac matrices are anti-symmetric, two 
antisymmetric Lorentz spinor indices are equivalent to a vector index, while 
three antisymmetric lower Lorentz spinor indices are equivalent to a single raised Lorentz 
spinor index by making use of the Levi-Civita symbol $\eps^{\a\b\g\d}$. 
The $n = 2$ case is exceptional in the sense that the linear multiplet 
is a gauge multiplet because $G_a = \frac{1}{4} (\tilde{\g}_a)^{\a\b} G_{\a\b}$ 
may be identified with the dual of the field strength of a four-form gauge field 
since $G_a$ is divergenceless. For $n\geq3$ the supermultiplets  
do not possess similar restrictions at the component level and are therefore not gauge multiplets. 
\begin{table}
\begin{center}
\begin{tabular}{ |c|c|c| }
\hline
component field & dimension \\
\hline
$\varphi^{i_1 \cdots i_n}$ & $2 n$ \\ 
$\psi_\a{}^{i_1 \cdots i_{n-1}}$ & $ 2 n+{1\over2}$ \\ 
$G_{[\a\b]}{}^{i_1 \cdots i_{n - 2}}$ & $ 2 n+1$ \\ 
$\psi_{[\a\b\g]}{}^{i_1 \cdots i_{n - 3}}$ & $ 2 n+{3\over2}$ \ \\ 
$\varphi{}^{i_1 \cdots i_{n-4}}$ & $ 2 n+2$ \\ 
\hline
\end{tabular}
\caption{Field content of the $\cO(n)$ supermultiplet} \label{O(n)fieldContent}
\end{center}
\end{table}

Since we will be concerned with the cases for which prepotential formulations exist, we restrict ourselves to the 
class of multiplets with $n\geq2$. Indeed,  in addition to $n=2$, which was discussed above,  
prepotential formulations can also be given for $n \geq 3$. 
As $n=3$ and $n>3$ need separate treatment, we will discuss them in turn. 


\subsection{$n = 3$}

We can solve the defining constraint for the $\cO(3)$ multiplet in terms of an unconstrained 
prepotential as follows
\be {L}^{ijk} = \nabla^{ijkl} \nabla_{\a l} V^\a \ , 
\ee
where the prepotential $V^\a$ is defined up to the gauge transformations
\be 
\d V^\a = \nabla_{\b j} \z^{\a\b j} \ , \quad \z^{\a\b i} = \z^{(\a\b) i} \ .
\ee
One can check that $L^{ijk}$ is primary using the fact that $V^\a$ is a primary superfield 
of dimension $7/2$.


\subsection{$n \geq4$}

For these cases, the solution of \eqref{O(n)DefiningConstraint} is 
\be
{L}^{i_1 \cdots i_{n}} = \nabla^{(i_1 \cdots i_4} V^{i_5 \cdots i_{n})} \ , 
\label{O(p)prepotForm}
\ee
which is invariant under the gauge transformations of the unconstrained prepotential
\be 
\d V^{i_1 \dots i_{n-4}} = \nabla_\a^{(i_1} \z^{\a i_2 \cdots i_{n - 4})} \ .
\ee
The $\cO(n)$ multiplet \eqref{O(p)prepotForm} is primary. This is a consequence  
of the fact that the 
prepotential $V^{i_1 \cdots i_{n-4}}$ is a primary superfield of dimension $2 n - 2$.
The analogue of the above prepotential formulation in 5D appeared in 
appendix G of \cite{BKNT-M5D}.


\section{New superconformal multiplets and their gauge prepotentials}

We will now introduce new multiplets that generalise the vector multiplet. 
To do this we focus on a key property of the vector multiplet, which is that an invariant 
can be constructed by multiplying its prepotential with an $\cO(2)$ multiplet as in eq. \eqref{BFinv}. 
This implies that the index structure of the prepotential for the vector multiplet is such that  
it can be contracted with $L^{ij}$ to form an $SU(2)$ singlet. In generalising this property 
we seek to find multiplets that have prepotentials of the form $U_{i_1 \cdots i_n}$ 
that can be used to build a gauge invariant expression with an $\cO(n)$ multiplet as 
an integral over full superspace. In this way 
the new multiplets will be `dual' to the $\cO(n)$ multiplets.

\begin{table}
\begin{center}
\begin{tabular}{ |c|c|c| } 
\hline
component field & dimension \\
\hline
$V_{[\a\b]}$ & $1$ \\ 
$\l^{\a i}$ & $3/2$ \ \\ 
$Y^{ij}$ & $2$ \\ 
\hline
\end{tabular}
\caption{Field content of the vector multiplet}
\label{VMfieldContent}
\end{center}
\end{table}
The field content of the vector multiplet is summarised in Table \ref{VMfieldContent}, where 
$V_a = \frac{1}{4} (\tilde{\g}_a)^{\a\b} V_{\a\b}$ is a gauge field. Comparison with 
Table \ref{O(n)fieldContent} shows that every component field of the 
vector multiplet can be contracted with a component field of the $\cO(2)$ 
multiplet to give a scalar of dimension 6. The analogous property also 
holds for the component fields of the new multiplets. We have to distinguish three cases, depending 
on the number of $SU(2)$ indices of the prepotential. 
Note that 
while the vector multiplet is a gauge multiplet, the new multiplets are not.


\subsection{$n = 3$}

Starting from a 
primary superfield $\Tmult_\a$ of dimension $-3/2$ we impose the differential constraint
\be \nabla_{(\a}^i \Tmult_{\b)}^{\phantom{j}} = 0 \ .
\ee
This constraint can be solved in terms of a primary dimension $-4$ unconstrained prepotential $U_{ijk}$ 
as follows
\be \Tmult_\a = \nabla_{\a l} \nabla^{ijkl} U_{ijk} \ ,
\ee
where the prepotential is defined up to the gauge transformations
\be \d U_{ijk} = \nabla_\a^l \xi^\a{}_{ijkl} 
\ .
\ee
The field content of this new multiplet is summarised in Table \ref{XialphafieldContent}.

\begin{table}
\begin{center}
\begin{tabular}{ |c|c|c| } 
\hline
component field & dimension \\
\hline
$\l_\a$ & $- 3/2$ \\ 
$V_{[\a\b]}{}^i$ & $- 1$ \\ 
$\l_{[\a\b\g]}{}^{ij}$ & $-1/2$ \\ 
$Y^{ijk}$ & $0$ \\ 
\hline
\end{tabular}
\caption{Field content of the supermultiplet described by $\Tmult_\a$}
\label{XialphafieldContent}
\end{center}
\end{table}


\subsection{$n = 4$}

We can define another supermultiplet by a primary 
dimension $-4$ scalar superfield $\Tmult$ satisfying 
the constraint
\be \nabla^k_{(\a} \nabla_{\b) k}^{\phantom{k}} \Tmult = 0 \implies \nabla_{(\a}^i \nabla_{\b)}^j \Tmult = 0 \ .
\ee
One can solve this in terms of an unconstrained prepotential 
$U_{ijkl}$ 
as follows\footnote{This multiplet and its prepotential 
description first appeared in Minkowski superspace in \cite{HST83}.}
\be \Tmult = \nabla^{ijkl} U_{ijkl} \ , 
\ee
where $U_{ijkl}$ is primary of dimension $-6$. It is straightforward to check 
that $\Tmult$ is primary given $U_{ijkl}$ is primary. 
One can also check that $U_{ijkl}$ is defined up to the gauge transformations
\be \d U_{ijkl} = \nabla_\a^m \xi^\a{}_{ijklm} \ . 
\label{UGTofXi}
\ee
To check the invariance of $\Tmult$, the following identities are useful:
\be \nabla_\a^{(m} \nabla^{ijkl)} = \nabla^{(ijkl} \nabla_\a^{m)} = \nabla^m_{(\a} \nabla_{\b) m}^{\phantom{m}} \nabla^{ijkl} 
= \nabla^{ijkl} \nabla^m_{(\a} \nabla_{\b) m}^{\phantom{m}} = 0 \ .
\ee
The field content of this multiplet is summarised in Table \ref{XiscalarfieldContent} with $n = 4$.
\begin{table}
\begin{center}
\begin{tabular}{ |c|c|c|c|c| } 
\hline
component field & dimension \\
\hline
$\phi^{i_1 \cdots i_{n-4}}$ & $ 4 - 2 n$ \\ 
$\l_\a{}^{i_1 \cdots i_{n - 3}}$ & $ {9\over2} - 2 n$ \\ 
$V_{[\a\b]}{}^{i_1 \cdots i_{n - 2}}$ & $ 5 - 2 n $ \\ 
$\l_{[\a\b\g]}{}^{i_1 \cdots i_{n-1}}$ & ${11\over2} - 2 n $ \\ 
$Y^{i_1 \cdots i_{n}}$ & $6 - 2 n$ \\ 
\hline
\end{tabular}
\caption{Field content of the $\cO{}^{*}(n)$ multiplet}
\label{XiscalarfieldContent}
\end{center}
\end{table}


\subsection{$n \geq 5$}

As a generalisation of the scalar superfield we used in the previous subsection, we
can introduce superfields $\Tmult_{i_1 \cdots i_{n-4}}$ 
of dimension $4-2n$ 
satisfying the 
constraint
\be \nabla_\a^j \Tmult_{i_1 \cdots i_{n-5} j}^{\phantom{j}} = 0 \quad \implies 
\quad 
\nabla_{(\a}^j \nabla_{\b) j}^{\phantom{j}} \Tmult_{i_1 \cdots i_{n-4}} = 0 \ . \label{transverseConst}
\ee
Its solution in terms of an unconstrained prepotential is
\be \Tmult_{i_1 \cdots i_{n-4}} = \nabla^{i_{n-3} \cdots i_{n}} U_{i_1 \cdots i_{n}} \ ,
\ee
where $U_{i_1 \cdots i_{n}}$ is defined up to the gauge transformations
\be 
\d U_{i_1 \cdots i_{n}} = \nabla_{\a}^j \xi^{\a}{}_{i_1 \cdots i_{n} j} 
\ .
\label{3.10}
\ee
The field content of these multiplets is summarized in Table \ref{XiscalarfieldContent}. The form of 
the constraints \eqref{transverseConst} is `transverse-like' as opposed to the 
`longitudinal-like' constraints \eqref{O(n)DefiningConstraint}
of the $\cO(n)$ multiplets and for this reason we may refer to 
the new multiplets as transverse multiplets. Although in this 
sense it might be natural to refer to the $\cO(n)$ multiplets as longitudinal multiplets,  
we will 
stick to the well-established nomenclature for them. As for the transverse multiplets, we will 
denote them by $\cO{}^{*}(n)$ as a reminder that they were constructed to 
be `dual' to the $\cO(n)$ multiplets in the sense that one will be able 
to use them to construct an invariant as a product of it with an $\cO(n)$ multiplet. 
For convenience, we also define $\cO{}^{*}(2)$ to be the vector multiplet, 
$\cO{}^{*}(3)$ to be the multiplet described by the superfield $\Tmult_\a$ and $\cO{}^{*}(4)$ to be the multiplet 
described by the scalar superfield $\Tmult$. We have not defined $\cO{}^{*}(1)$ here but we will  
introduce it in section \ref{conclusion}.


\section{New superconformal invariants}

We will now  use the new multiplets and their prepotentials 
to generalise the $BF$ invariant \eqref{BFinv}.


\subsection{Generalised $BF$ invariants}

With the help of the results of the previous sections, one 
can immediately write down a supersymmetric invariant
\be I_{(n)} = \int \rd^{6|8} z\, E \, U_{i_1 \cdots i_n} \, {L}^{i_1 \cdots i_n} \ , \label{genBFinv}
\ee
where $E = {\rm Ber}(E_M{}^A)$ and, as usual,  $E^A = \rd z^M E_M{}^A$
is the supervielbein.
Taking into account the differential constraint on the $\cO(n)$ multiplet, it is straightforward to 
check that $I_{(n)}$ is invariant with respect to the gauge transformations of the prepotential $U_{i_1 \cdots i_n}$.
In the $n = 2$ case the invariant \eqref{genBFinv} corresponds to the standard $BF$ 
invariant \eqref{BFinv}, while for $n \geq 3$ we 
obtain new generalisations.\footnote{The invariant $I_{(4)}$ is similar to the one used in the 
description of the relaxed hypermultiplet action \cite{HST83} in Minkowski superspace.}
In the flat case, the supersymmetric invariant \eqref{genBFinv} 
can be seen to naturally originate in harmonic superspace, see Appendix A.

Note that one can equivalently define the generalised $BF$ invariants by explicitly using the 
prepotential for the $\cO(n)$ multiplets. For $n = 3$ eq. \eqref{genBFinv} is equivalent to
\be I_{(3)} = \int \rd^{6|8} z \, E \, V_\a \Tmult^\a
\ ,
\ee
while for $n \geq 4$ it is equivalent to
\be I_{(n)} = \int \rd^{6|8} z \, E \, V_{i_1 \cdots i_{n-4}} \Tmult^{i_1 \cdots i_{{n-4}}} \label{altgenBFinv}
\ .
\ee
However, as one can see from the above, the fact that the prepotential obtains a Lorentz index 
for the $n \leq 3$ cases means that we cannot treat 
all cases ubiquitously in this form.


\subsection{Superform realisation of the invariants}

It is worth mentioning that a {\it manifestly gauge-invariant} form of the invariant \eqref{genBFinv} for $n \geq 3$
can be given by making use of the superform construction in \cite{ALR16, BKNT}. It is 
based on the existence of a closed six-form in superspace that describes a locally superconformal invariant. 
It is expressed entirely in terms of a basic primary superfield  $A_\a{}^{ijk}$ of dimension 9/2 which satisfies 
the differential constraint $\nabla_{(\a}^{(i} A_{\b)}^{\phantom{j}}{}^{jkl)} = 0$. The
superfield $A_\a{}^{ijk}$ plays the role of a supersymmetric Lagrangian similar to the 
chiral Lagrangian in 4D. Its component structure was elaborated in \cite{BNT-M17}. 

The invariant \eqref{genBFinv} can be equivalently described by a composite 
superfield $A_\a{}^{ijk}$ which is, roughly, the product of an $\cO{}^{*}(n)$ multiplet 
and the corresponding $\cO(n)$ multiplet.
More specifically, for $n = 3$ one has
\be A_\a{}^{ijk} \propto \Tmult_\a {L}^{ijk} \ ,
\ee
while for $n \geq 4$, one takes
\be A_\a{}^{ijk} \propto (1 + n) \Tmult_{i_1 \cdots i_{n- 4}} \nabla_{\a l} {L}^{ijkl i_1 \cdots i_{n -4}}
+ (5 + n) (\nabla_{\a l} \Tmult_{i_1 \cdots i_{n-4}}) {L}^{ijkl i_1 \cdots i_{n -4}} \ . 
\label{gennCompositeA}
\ee
The above representations are democratic in the sense that they 
put both multiplets on the same footing.\footnote{Any full superspace 
invariant $\int \rd^{6|8} z\, E \, \cL$, with the Lagrangian $\cL$ being a primary scalar superfield of dimension $+2$,  
can equivalently be described by \eqref{gennCompositeA} 
with a composite $\cO(4)$ multiplet ${\mathbb L}^{ijkl} = \nabla^{ijkl} (\Tmult^{-1} \cL)$, where 
$\Tmult$ is chosen to be nowhere vanishing. Its dependence on 
$\Tmult$ is artificial since the invariant is independent of $\Tmult$ modulo a total derivative.}

The above results are useful in elaborating the component structure 
since they make the connection with the component action more direct through the 
results of \cite{BKNT, BNT-M17}. Note, however, that while a manifestly gauge-invariant description 
exists for $n \geq 3$, an analogous description in terms of 
the superfield $A_\a{}^{ijk}$ does not exist for $n = 2$.


\section{Higher derivative superconformal invariants}

With the results obtained so far, it is possible to construct composite $\cO(n)$ 
multiplets and use them to construct higher derivative actions for the 
$\cO{}^{*}(n)$ multiplets. In this section we will restrict ourselves to a 
superconformally flat geometry, where the super-Weyl tensor vanishes, $W^{\a\b} = 0$.

One can construct a composite $\cO(3)$ multiplet out 
of the superfield $\Tmult_\a$ as follows
\be {\bm L}^{ijk} = \Box^3 \nabla^{\a ijk} \Tmult_\a  \ . \label{n=3Compositehder}
\ee
The power of the d'Alembertian operator, $\Box := \nabla^a \nabla_a$, is chosen to ensure that 
${\bm L}^{ijk}$ has the right dimension and is primary in the superconformally flat geometry. 
To check that it is primary, the following identities are useful:
\bsubeq \label{usefulIdentitiesSnab}
\bea
\left[ S^\a_i , \Box \right] &=& - 2 \ri \nabla^{\a\b} \nabla_{\b i} + \ri [\nabla^{\a\b} , \nabla_{\b i}] \ , \\
\{ S^\a_l , \nabla^{\b ijk} \} &=& 8 \d_l^{(i} \nabla^{\a\b jk)} + \nabla^{\b\g (ij} \{ S^\a_l , \nabla_\g^{l)} \} 
\ .
\eea
\esubeq
A higher derivative 
invariant can now be described by plugging in ${\overline{\bm L}}^{ijk}$
into the generalised $BF$ invariant \eqref{genBFinv}, giving
\be J_{(3)} = \int \rd^{6|8} z \, E \, U_{ijk} {\overline{\bm L}}^{ijk} \ . \label{highern=3action}
\ee
This invariant is the supersymmetric extension of a $\overline{\l}_\a \Box^4 \partial^{\a\b} \l_\b$ term.

We can also 
describe higher derivative invariants for the $\cO{}^{*}(n)$ multiplets with $n \geq 4$. We begin by 
building a composite $\cO(n)$ multiplet from the superfield $\Tmult_{i_1 \cdots i_{n-4}}$:
\be {\bm L}^{i_1 \cdots i_{n}} = \Box^{2n-3} \nabla^{(i_1 \cdots i_4} \Tmult^{i_5 \cdots i_{n})} \ . \label{genCompositehder} 
\ee
One can show that in a superconformally 
flat geometry ${\bm L}^{i_1 \cdots i_{n}}$ is primary, 
using again eqs. \eqref{usefulIdentitiesSnab} and the following identities:
\bsubeq
\bea
\left[ S^\a_m , \nabla^{ijkl} \right] &=& 3 \d_m^{(i} \nabla^{\a jkl)} - \frac{1}{4} \nabla^{\b (ijk} \{ S^\a_m , \nabla_\b^{l)} \} \ , \\
\nabla^{\a\b} \nabla_{\b\g} &=& - \d_\g^\a \Box + \hf [\nabla^{\a\b} , \nabla_{\b\g}] \ , \\
\nabla_{\a j} \nabla^{(ji_1 i_2 i_3} \Tmult^{i_4 \cdots i_{n-1})}
&=& - \frac{2 (n+1) \ri}{n} \nabla_{\b\g} \nabla^{\g (i_1 i_2 i_3 } \Tmult^{i_4 \cdots i_{n-1})}
+ ({W^{\a\b} \ {\rm terms}}) \ .
\eea
\esubeq
The higher derivative invariants are now described by the generalised 
$BF$ invariant \eqref{genBFinv} with $\overline{\bm L}^{i_1 \cdots i_n}$, i.e.
\be J_{(n)} = \int \rd^{6|8} z \, E \, U_{i_1 \cdots i_{n}} {\overline{\bm L}}^{i_1 \cdots i_{n}} 
= \int \rd^{6|8} z \, E \, T_{i_1 \cdots i_n} \Box^{2 n - 3} {\overline{T}}^{i_1 \cdots i_n}
\ . \label{highernqeq4action}
\ee
$J_{(n)}$ contains a term of the form $\overline{\phi}^{i_1 \cdots i_{n-4}} \Box^{2 n - 1} \phi_{i_1 \cdots i_{n-4}}$ 
at the component level.


\section{Discussion} \label{conclusion}

A family of higher derivative superconformal actions was proposed 
in \cite{Beccaria:2015ypa, Beccaria:2017dmw}. Upon appropriate identification 
of the fields, one can see that the actions sketched in \cite{Beccaria:2015ypa, Beccaria:2017dmw} 
correspond to the higher derivative superconformal invariants  
presented in this paper. Note that analogous higher derivative invariants realised in terms of $\cO(n)$ 
multiplets is not possible since all its fields have dimension $2 n$ or higher. 
As a result, the only case when a local 
superconformal action, that is quadratic in the fields, could exist is $n=1$, but this 
corresponds to an on-shell hypermultiplet. This tells us that such higher derivative invariants 
require the use of the $\cO{}^{*}(n)$ multiplets.


The question whether the higher derivative invariants of the previous section 
can be lifted to conformal supergravity is a separate problem. As a matter of fact, it appears not to be the case. 
The point is that \eqref{highernqeq4action} should include the term $\bar{\phi} \Box^{2n - 1} \phi$ once one truncates
to $\cN = 0$ and it is known that there does not exist a locally conformal completion for $n \geq 3$ \cite{Gover:2003am}. 
The same is expected for the higher derivative invariant \eqref{highern=3action}.
In conformally flat backgrounds the obstruction for their existence is, however, absent
\cite{Juhl}.

The results in this paper may have interesting consequences in the construction of new higher derivative supergravity 
theories. To construct Poincar\'e supergravity theories within a superconformal framework, one must 
use a conformal compensating multiplet, which is a real scalar superfield $\Phi$ that we choose, 
without loss of generality, to have dimension one. 
In this framework it becomes apparent that the construction of composite $\cO(n)$ multiplets from the  $\cO{}^{*}(n)$ multiplets 
can be made in many more ways using the results of this paper. For instance, we can construct $\cO(n)$ multiplets 
with $n \geq 4$ as follows
\be {\bm L}^{i_1 \cdots i_{n}} = \nabla^{(i_1 \cdots i_4} (\Tmult^{i_5 \cdots i_{n})} \Phi^{4 n - 6} ) \ . \label{comps1}
\ee
Furthermore, we can choose the compensator as $\Phi^{-4(n-2)} = \Tmult^{i_1 \cdots i_{n-4}} \overline{\Tmult}_{i_1 \cdots i_{n-4}}$ 
leading to a four-derivative invariant for the $\cO{}^{*}(n)$ multiplet 
once one plugs the composite into the generalised $BF$ invariant \eqref{genBFinv}.
We will explore the supercurrents of 
such theories in a forthcoming paper \cite{KNTtba}.

It is interesting to note that supersymmetric invariants built
 out of $\cO{}^{*}(n)$ for $n \geq 4$ may be mapped to invariants of the $\cO(2)$ multiplet 
 for even $n = 2 m$. One only needs to make use of an $\cO(2)$ multiplet $L^{ij}$ 
such that $L^2 := L^{ij} L_{ij}$ is nowhere vanishing. Given the $\cO(2)$ multiplet, we can
choose the $\cO{}^{*}(2m)$ multiplet to be\footnote{It is a straightforward 
exercise to check that the composite multiplet satisfies the appropriate differential constraint.}
\be {\bm T}^{i_1 \cdots i_{2m}} = L^{- (2m +1)} L^{(i_1i_2} \cdots L^{i_{2m-1} i_{2m})} \ .
\ee
In this way the resulting invariant will be expressed entirely in terms of the $\cO(2)$ multiplet.

It is also possible to construct 
composite $\cO{}^{*}(n)$ multiplets out of $\cO(n)$ multiplets leading 
to superconformal invariants for the $\cO(n)$ multiplets. 
For instance, for $n \geq 4$ we can take the composite
\be \bm \Tmult_{i_1 \cdots i_{n-4}} = \nabla^{i_{n-3} \cdots i_{n}} (L_{i_1 \cdots i_{n}} \Phi^{- 4 n + 2}) \ . \label{comps2}
\ee
We can choose the compensator to be $\Phi^{4 n} = L^{i_1 \cdots i_{n}} \overline{L}_{i_1 \cdots i_{n}}$ and 
then plugging the result into the invariant \eqref{altgenBFinv} gives a four-derivative invariant for 
the $\cO(n)$ multiplet. Many more invariants can be constructed by iterating eqs. \eqref{comps1} 
and \eqref{comps2} in various ways, and using the fact that there is an obvious 
multiplication defined on the space of $\cO(n)$ multiplets and the space of $\cO{}^{*}(n)$ multiplets. 
Similar constructions appeared in 4D for the $\cO(2)$ multiplet in \cite{BK11}.

In this paper, we have defined the $\cO{}^{*}(n)$ multipelts for $n \geq 2$. The $\cO{}^{*}(n)$ multiplet has the same 
number of degrees of freedom as the corresponding $\cO(n)$ multiplet as follows from the generalised 
$BF$ invaraint \eqref{genBFinv}. We have ignored the $n = 1$ case since the $\cO(1)$ multiplet is 
on-shell and does not have a prepotential formulation. However, it is interesting to ask whether there exists 
another on-shell multiplet possessing the same number of degrees of freedom as that of the $\cO(1)$ multiplet that 
one could naturally add to the class of $\cO{}^{*}(n)$ multiplets. Such a superconformal 
multiplet is described by a primary superfield 
$\Tmult^\a$ of dimension $- 1/2$ satisfying the constraint
\be \nabla_\a^i \Tmult^\b = \frac{1}{4} \d_\a^\b \nabla_\g^i \Tmult^\g \ .
\ee
It might be appropriate to define this multiplet as the $\cO{}^{*}(1)$ multiplet. 

The multiplets described in this paper certainly do not exhaust all 
possible superconformal multiplets nor those that permit a prepotential formulation. 
To illustrate this we introduce a supermultiplet described by a 
superfield $M_{\a_1 \cdots \a_n} = M_{(\a_1 \cdots \a_n)}$ with $n \geq 1$ 
and subject to the constraint\footnote{These superconformal constraints 
appeared in Minkowski superspace in \cite{Park}.}
\be \nabla_{(\a_1}^i M_{\a_2 \cdots \a_{n+1})} = 0 \ ,
\ee
which describes a superconformal multiplet when 
$\mathbb D M_{\a_1 \cdots \a_n} = - \frac{3 n}{2} M_{\a_1 \cdots \a_n}$. For 
$n > 2$ the differential constraint can be solved in terms of a prepotential as
\be M_{\a_1 \cdots \a_n} = \nabla_{(\a_1}^k \nabla_{\a_2 k} V_{\a_3 \cdots \a_n)} \ ,
\ee
where the prepotential $V_{\a_1 \cdots \a_{n-2}} = V_{(\a_1 \cdots \a_{n-2})}$ possesses the gauge transformations
\be \d V_{\a_1 \cdots \a_{n-2}} = \nabla_{(\a_1}^k \xi_{\a_2 \cdots \a_{n-2}) k} \ .
\ee
Interestingly, upon introducing a `dual' multiplet described by a superfield $N^{\a_1 \cdots \a_{n-2}} = N^{(\a_1 \cdots \a_{n-2})}$ 
that is constrained by $\nabla_\b^k N^{\a_1 \cdots \a_{n-3} \b} = 0$, 
one can write down a gauge invariant
\be
\int \rd z^{6|8} \, E \, N^{\a_1 \cdots \a_{n-2}} V_{\a_1 \cdots \a_{n-2}} 
\ .
\ee
However, it does not appear that such an invariant can be used to describe higher derivative superconformal actions.

It is worthwhile mentioning that the new multiplets and constructions introduced in this paper imply the existence of 
analogues in lower spacetime dimensions. We leave their exploration for future work.

$~$\\
\noindent
{\bf Acknowledgements:}\\
SMK is grateful to Arkady Tseytlin for discussions. 
The work of SMK was supported in part by the Australian 
Research Council,  project No. DP160103633.
JN and ST acknowledges support from GIF -- the German-Israeli Foundation 
for Scientific Research and Development.


\appendix

\section{Harmonic superspace construction}

In this appendix we present a harmonic superspace origin of 
the supersymmetric invariant \eqref{genBFinv} in the flat case. 
Our construction is an extension of the known procedure to read off
the Mezincescu prepotential from the analytic prepotential 
  \cite{GIKOS,GIOS}.

Given an $\cO(n)$ multiplet $L^{i_1 \dots i_n} (z)$ in Minkowski superspace, 
with $ n \geq 2$,
\bea
D_\a^{(i_1} L^{i_2 \cdots i_{n+1})} = 0~ , 
\eea
we associate with it a harmonic superfield 
\bea
L^{(+n)} (z, u) = L^{i_1 \dots i_n} (z) u^+_{i_1} \dots u^+_{i_n}~, 
\eea
which is analytic 
\begin{subequations}
\bea
D^+_\a L^{(+n)} = 0~, \qquad
D^\pm_\a:= u^\pm_i {D}^i_\alpha
\eea
and obeys the harmonic shortness constraint 
\bea
{D}^{++} L^{(+n)} = 0~.
\eea
\end{subequations}
Then we can define a supersymmetric invariant 
\bea
\hat I_{(n)} =
 \int {\rm d}\zeta^{(-4)}\, \O^{(4-n)} L^{(+n)}~, \qquad D^+_\a \O^{(4-n)} =0~,
 \label{A.4}
 \eea
in which the integrand involves an analytic potential $ \O^{(4-n)} $.
The integration in \eqref{A.4}
is carried out over the analytic subspace of the harmonic superspace. 
It is defined by 
\bea
 \int {\rm d}\zeta^{(-4)}\,  \cL^{(4)} &=& \int \rd^6 x \int \rd u \, 
  ({D}^-)^4 \cL^{(4)}~, \qquad  D^+_\a \cL^{(4)} =0~, \non \\
 && (D^-)^4:= -\frac1{96} \varepsilon^{\alpha\beta\gamma\delta}
{D}^-_\alpha {D}^-_\beta {D}^-_\gamma
{D}^-_\delta~,
\eea
for any analytic Lagrangian $\cL^{(4)} $.
 Here the $u$-integral denotes the integration over the group manifold $\rm SU(2)$ defined as in  \cite{GIKOS,GIOS}. 
The functional \eqref{A.4} is invariant under gauge transformations of the form 
\bea
\d  \O^{(4-n)} = D^{++}  \L^{(2-n)}~, \qquad D^+_\a  \L^{(2-n)}=0~,
\label{A.5}
\eea
with the gauge parameter $ \L^{(2-n)}$ being an analytic superfield of U(1) 
charge $(2-n)$. The harmonic derivative $D^{++}$ is defined as usual
\cite{GIKOS,GIOS}.

The analyticity constraint on $ \O^{(4-n)}$ is solved by 
\bea
 \O^{(4-n)} =   ({D}^+)^4 U^{(-n)}~, \qquad
 (D^+)^4:= -\frac1{96} \varepsilon^{\alpha\beta\gamma\delta}
{D}^+_\alpha {D}^+_\beta {D}^+_\gamma
{D}^+_\delta~.
 \label{A.6}
 \eea
 Here the prepotential $U^{(-n)}(z,u)$ is an unconstrained harmonic superfield
 which is defined modulo gauge transformations 
 \bea
 \d U^{(-n)} = D^+_\a \x^{(-n-1) \, \a}~,
\label{A.7}
 \eea
 with the gauge parameter $\x^{(-n-1)\,  \a} (z,u)$ being an unconstrained 
 harmonic superfield.
 
The analyticity constraint on the gauge parameter in \eqref{A.5} can be solved 
similarly to the representation \eqref{A.6} for $ \O^{(4-n)}$, 
\bea
 \L^{(2-n)} =   ({D}^+)^4 \r^{(-n-2)}~. 
 \eea
 Then the gauge transformation becomes equivalent to 
 \bea
  \d U^{(-n)} = D^{++}  \r^{(-n-2)} ~.
  \label{A.9}
  \eea
  The unconstrained harmonic superfields $U^{(-n)} $ and $\r^{(-n-2)} $
  can be represented by convergent harmonic series 
\begin{subequations}
\bea
U^{(-n)}(z,u) &=& U^{i_1 \dots i_n} (z)u^-_{i_1} \dots u^-_{i_n} \non \\
&& +\sum_{m=1}^\infty U^{(i_1 \dots  i_{n+2m})} (z)
u^+_{i_1} \dots u^+_{i_{m}} u^-_{i_{m+1}} \dots u^-_{i_{n+2m}} ~, \\
\r^{(-n-2)}(z,u) &=&
\sum_{m=0}^\infty \r^{(i_1 \dots  i_{n+2+2m})} (z)
u^+_{i_1} \dots u^+_{i_{m}} u^-_{i_{m+1}} \dots u^-_{i_{n+2+2m}} ~.
\eea
\end{subequations}
It follows from these expressions that the gauge symmetry \eqref{A.9} allows us 
to impose a gauge condition
\bea
U^{(-n)}(z,u) &=& U^{i_1 \dots i_n} (z)u^-_{i_1} \dots u^-_{i_n} ~,
\label{A.11}
\eea
which completely fixes the gauge freedom \eqref{A.9}.
In this gauge, we still have the freedom to perform 
those gauge transformations \eqref{A.7} which are 
generated by parameters of the form\footnote{Such a transformation 
should be  accompanied by a compensating $\r$-transformation \eqref{A.9}
which is required to preserve the gauge condition \eqref{A.11}.}
\bea
\x^{(-n-1)\,  \a} (z,u) 
= \x^{i_1 \dots i_{n+1}\, \a} (z)u^-_{i_1} \dots u^-_{i_{n+1}}  ~.
\eea
The resulting gauge transformation of the prepotential $U^{i_1 \dots i_n} (z)$
coincides with the flat-superspace version of 
\eqref{3.10}.

In the gauge \eqref{A.11}, the supersymmetric invariant \eqref{A.4} can be 
represented as an integral over Minkowski superspace by making use of  
the identity 
\bea
 \int {\rm d}\zeta^{(-4)}\,
  ({D}^+)^4 \cL(z,u)
=   \int {\rm d}^{6|8}z \int  {\rm d}u\, \cL(z,u) 
  ~,
\eea
and computing the relevant harmonic integral. This reduces the supersymmetric 
invariant \eqref{A.4} to the flat-superspace version of \eqref{genBFinv} modulo a numerical factor. 

In the $n=2$ case, the supersymmetric invariant \eqref{A.4} is the harmonic superspace 
realisation of the linear multiplet action.


\begin{footnotesize}

\end{footnotesize}

\end{document}